# Structural effect on phonon attenuation in metallic liquids and glasses

Jaeyun Moon[1*] and Takeshi Egami[1,2,3†]

[1]*Materials Science and Technology Division,
Oak Ridge National Laboratory, Oak Ridge, Tennessee 37831, USA*
[2]*Department of Materials Science and Engineering,
University of Tennessee, Knoxville, Tennessee 37996, USA*
[3]*Department of Physics and Astronomy,
University of Tennessee, Knoxville, Tennessee 37996, USA*

\* *Electronic mail*: moonj@ornl.gov , †Electronic mail: egami@utk.edu

**Abstract**

The attenuation rate of vibrational excitations in various metallic liquids and glasses has been reported to change from the quadratic dependence on wavevector at low wavevectors to the linear dependence at high wavevectors. However, the origin of this behavior is not clear. Here, the analysis of this phenomenon through molecular dynamics is presented for prototypical metallic liquids, $Cu_{56}Zr_{44}$ and Fe. It is shown that the crossover wavevector is strongly correlated with the structural coherence length characterizing coarse-grained density correlations. We suggest that the linear dependence is caused by scattering of vibrational excitations by structural activation processes with low activation energies which are distinctively observed in metallic systems.

Understanding the nature of vibrational excitations is key to elucidating dynamical and thermodynamic properties of condensed matters, such as thermal conductivity and heat capacity. Whereas the vibrational excitations in crystals are well-described as phonons [1,2], those in liquids and glasses are more difficult to characterize due to dynamic disorder and lack of structural periodicity [3–8]. Consequently, they are investigated principally by experimental studies including inelastic scattering measurements [9–16] and picosecond acoustics [17–19], and by molecular dynamics simulations of dynamic structure factor [20–23].

In crystalline solids anharmonicity plays a major role in phonon scattering. However, prior inelastic scattering measurements and simulations of dynamic structure factor on numerous non-metallic glasses have demonstrated that anharmonicity is important only at very low wavevectors of the order of 0.01 nm$^{-1}$ [24,25]. At higher wavevectors, up to slightly below 1 to 2 nm$^{-1}$, the attenuation rate of longitudinal vibrational excitations displays quartic power law dependence on wavevector, which has been ascribed to Rayleigh scattering [26–28]. This is followed by a quadratic power law at higher wavevectors [9,24,26]. The quadratic power law has also been observed in liquids [9]. There have been intense research efforts in the past few decades on the origin of the quadratic power law. Some attribute it to the boson peak and/or Ioffe-Regel crossover in this wavevector region [29,30]. Several theories and models have been proposed, linking it to random spatial fluctuations of transverse elastic constants [31] and diffusons, which are non-propagating but non-localized normal modes [32]. These efforts attained some degrees of success, but consensus is yet to be reached.

Interestingly, for various metallic liquids and glasses an additional crossover behavior from quadratic to linear dependence has been observed at a few nm$^{-1}$ by measurements [33–35] and simulations [36,37], associated with some unknown dynamical effects [33]. Because the crossover frequencies and wavevectors of Ioffe-Regel localization are typically higher for metallic systems than for other network and polymeric systems and lie in this linear regime [38], it is of major interest from both fundamental and practical perspectives to understand the nature of vibrational excitations in this range of wavevectors and identify their scattering mechanisms.

In this Letter, we investigate the origin of the linear power dependence in the attenuation rates of longitudinal vibrational excitations for liquid $Cu_{56}Zr_{44}$ and Fe, prototypical simple metallic systems, through molecular dynamics. By looking at the temperature dependence over a wide range of temperature, we observe clear evidence of strong correlation between the crossover wavevector from quadratic to linear power law and the structural coherence lengths of atomic density-density correlations, providing unique insights into the origin of the linear attenuation rates in metallic liquids and glasses.

Molecular dynamics simulations were performed using the large scale massively parallel simulator (LAMMPS) [39] with a time step of 0.5 fs for $Cu_{56}Zr_{44}$ and 1 fs for Fe. Periodic boundary conditions were imposed. The EAM potential [40] was used for $Cu_{56}Zr_{44}$ (50,000 atoms) whereas the modified Johnson potential [41] was employed for Fe (64,000 atoms). To achieve fine resolutions in wavevectors ($k_z = \frac{2\pi n}{L_z}$), cuboid domains with a long side length along z-direction were utilized (2.5 nm by 2.5 nm by 136.4 nm for $Cu_{56}Zr_{44}$ and 2.4 nm by 2.4 nm by 147 nm for Fe). Similar elongated cuboid domains were used in prior studies [37] to characterize vibrational excitations. Both systems were heated to 5000 K to form liquids and were cooled at 100 K ps$^{-1}$ by 500 K increments down to 2000 K which is well above the glass transition temperatures of 700 K for $Cu_{56}Zr_{44}$ and 950 K for Fe. We used the NVT ensemble to prevent evaporation. At each temperature, the systems were equilibrated for 100 ps followed by data recording of atomic positions and velocities for 50 ps with 25 fs intervals. For better statistics, 5 independent runs with different initial velocities were carried out.

The longitudinal current correlation function, $C_L(k,\omega)$, has been widely used to characterize longitudinal vibrational excitations in numerous liquids and glasses [42,43] and is given by

$$C_L(k,\omega) = \frac{1}{2\pi N}\int dt \left\langle \sum_i^N [\boldsymbol{v}_i(t)\cdot \hat{\boldsymbol{k}}]\hat{\boldsymbol{k}} e^{-i\boldsymbol{k}\cdot \boldsymbol{r}_i(t)} \sum_i^N [\boldsymbol{v}_i(t)\cdot \hat{\boldsymbol{k}}]\hat{\boldsymbol{k}} e^{i\boldsymbol{k}\cdot \boldsymbol{r}_i(t)} \right\rangle e^{i\omega t} \quad (1)$$

where $\boldsymbol{v}_i(t)$ and $\boldsymbol{r}_i(t)$ are the time dependent velocity and position, respectively, of atom $i$ and the sum is over all atoms. The longitudinal current correlation function is related to the dynamic structure factor, $S(k,\omega)$, by, $C_L(k,\omega) = (\omega^2/k^2)S(k,\omega)$. The longitudinal current correlation functions for $Cu_{56}Zr_{44}$ and Fe at 5000 K are shown representatively in Fig 1 (a) and (b). For better visualization, each spectrum is normalized by its maximum for each wavevector. Well-defined dispersion curves are observed, especially at low wavevectors, followed by large broadening at high wavevectors as expected for disordered materials. To extract the vibrational excitation frequencies and attenuation rates accurately, each spectrum was Fourier transformed into time space to separate long-time noises which obscure the broadening in the frequency space. Each spectrum was then fitted by $F(t) = F(0)e^{-\Gamma t}\cos(2\pi f t)$ where $\Gamma$ is the attenuation rate and $f$ is the excitation frequency. We do not attempt to analyze the vibrational excitations with wavevectors below 1 nm$^{-1}$ because much longer data recording times are necessary for accurate analysis at low wavevectors where lifetimes are quite long. Instead, we focused on the crossover phenomenon which is observed at higher wavevectors at a few nm$^{-1}$. Typical time-resolved structure factors, $C_L(k,t)$, at small and large wavevectors are shown in Fig. 1 (c) and (d). The function $F(t)$ shows excellent fitting to the data, yielding excitation frequencies and attenuation rates for all temperatures for both systems.

Extracted wavevector dependent attenuation rates at 5000 K and 2000 K for liquid $Cu_{56}Zr_{44}$ and Fe are depicted in Fig. 2 (see Fig. S1 in Supplementary Material for data at all temperatures). For ease of comparison, the 2000 K data were multiplied by a factor of 0.1. A clear crossover from $\Gamma \propto k^2$ to $\Gamma \propto k$ is shown for both metallic systems. Interestingly, we find that the crossover wavevector, $k_{cross}$, varies with temperature. In both systems, $k_{cross}$ nearly halves (by a factor of 0.63 for $Cu_{56}Zr_{44}$ and 0.56 for Fe) going from 5000 K to 2000 K. We expect $k_{cross}$ saturate to a finite value near the glass transition, because the quadratic to linear crossover behavior is observed also in metallic glasses and attenuation rates at these wavevectors show very little temperature dependence for glasses.

To gain further insight into the crossover behavior of attenuation rates, we looked for other properties or structural features that could help explain the temperature dependence in the crossover wavevector. We found no correlation with dynamic properties such as the Maxwell relaxation time and the non-Gaussian parameter for displacements and structural properties such as the average coordination numbers, in terms of temperature dependence. However, direct correlation was found for the structural coherence lengths representing the medium-range order (MRO). The pair distribution function (PDF), $g(r)$, describes the distribution of interatomic distances and scales beyond the first peak as

$$g(r) \sim \frac{1}{r} e^{-\frac{r}{\xi_s}} \qquad (2)$$

where $\xi_s$ is the structural coherence length [44]. Unlike the first peak of $g(r)$ which describes the atom-atom (point-to-point) correlations with the nearest neighbor atoms, the MRO depicts the correlation between the central atom and the groups of atoms (point-to-set) representing coarse-grained density fluctuation [44,45]. The $\xi_s$ has been shown to be related to viscosity [46], liquid fragility [47], and elastic moduli [48].

The magnitude of the reduced PDF, $|G(r)|$, where $G(r) = 4\pi\rho_0 r[g(r) - 1]$ with $\rho_0$ being the atomic number density is plotted in Fig. 3 for $Cu_{56}Zr_{44}$ and Fe at 5000 K and 2000 K (for data at all temperatures, see Fig. S2 in Supplementary Material). The reduced

PDFs for 2000 K are scaled by two orders of magnitude for ease of comparison. A clear exponential decay is observed for all, from which the structural coherence length, $\xi_s$, can be determined. The $\xi_s$ follows the Curie-Weiss law in its temperature dependence [44] down to the glass transition temperature where it saturates to a finite value. The inverse of $\xi_s$ and the crossover wavevector are plotted for all temperatures for both systems in Fig. 4. We find clear resemblance in the temperature dependence of these two quantities across a wide range of temperatures in both systems, suggesting strong correlations between the linear regime of vibrational excitations attenuation rates and structural coherence lengths. This finding leads to important implications on the nature of vibrational excitations in metallic liquids and glasses.

In general, Matthiessen's rule applies to the overall lifetimes affected by several factors operating in parallel. For instance, 3-phonon process, 4-phonon process, boundary scattering and others will result in the total attenuation as

$$\frac{1}{\tau} = \Gamma = \Gamma_{3-phonon} + \Gamma_{4-phonon} + \Gamma_{boundary} + \cdots \quad (3)$$

According to this rule, we expect a linear dependence on *k* at low *k* and quadratic dependence above the crossover, because the linear term dominates at low *k* and the quadratic term is dominant at high *k* when extrapolated. However, actual observations are in total opposite, suggesting that the two mechanisms are not operating in parallel. Instead, a distinct mechanism is in operation in each domain of wavevector; when the phonon wavelength is shorter than $2\pi\xi_s$ a mechanism which produces the linear *k*-dependence is in operation, and when the wavelength is longer phonons are scattered by a different mechanism producing a quadratic dependence possibly by random spatial fluctuations of transverse elastic constants and diffusons as mentioned before. Interestingly, the defiance of Matthiessen's rule applies as well to the crossover from the Rayleigh-like scattering regime to the quadratic regime observed in many glasses [26].

There are only few scattering mechanisms that describe a linear power law of phonon attenuation rates reported in literature, including scattering by dislocations [49,50] and interactions with dipoles [51], both of which are irrelevant to metallic liquids and glasses. In disordered materials, potential energy landscapes and inherent structures obtained by removing kinetic energies are very useful concepts in interpreting complex phenomenology of atomic dynamics [52,53]. In our prior work [54], the potential landscapes of $Cu_{56}Zr_{44}$ was examined by studying the distribution of activation energies for the activation process from local minima to saddle points through Activation-Relaxation Technique (ART) [55]. It was found that metallic systems have a wide distribution of energy barriers with low activation energies below ~ 100 meV. These low barriers are not observed in other network glasses such as amorphous silicon [56]. We speculate that vibrational excitations activate atoms to overcome these low activation energy barriers through strain fields, resulting in energy loss. The strain fields are gradients of the vibrational displacements, so that the energy loss will be linear with wavevector. Furthermore, the activation energy for viscosity is known to be related to $\xi_s$ [43,45], suggesting a key to understanding the strong correlations between $k_{cross}$ and $\xi_s$. This picture can be considered as generalization of phenomenological two level system model [57–59] for liquids.

Our results provide unique insights into the attenuation rate crossover for longitudinal vibrational excitations from quadratic to linear dependence in wavevector in metallic liquids and glasses. By examining the crossover wavevector over a wide range of temperature, we observe strong relations between the crossover wavevector and the structural coherence lengths characterizing the density-density correlations. The crossover behavior is interpreted as the transition to the regime where vibrational excitations are scattered by strain field through structural activation over low energy barrier, which is distinctively observed in metallic systems. We believe that our work will stimulate future high temperature experiments to characterize the observed temperature dependence and theory development for the new scattering mechanism of vibrational excitations in disordered materials.

**Acknowledgment:**

The authors thank Prof. Frances Hellman and Dr. Lucas Lindsay for helpful discussions. This work was supported by the US Department of Energy, Office of Science, Basic Energy Sciences, Materials Sciences and Engineering Division. This work used the Extreme Science and Engineering Discovery Environment (XSEDE) Comet under allocation TG-MAT200012.

# References

[1] P. Debye, *Zur Theorie Der Spezifischen Wärmen*, Annalen Der Physik **344**, 789 (1912).

[2] A. Togo and I. Tanaka, *First Principles Phonon Calculations in Materials Science*, Scripta Materialia **108**, 1 (2015).

[3] P. B. Allen, J. L. Feldman, J. Fabian, and F. Wooten, *Diffusons, Locons and Propagons: Character of Atomie Yibrations in Amorphous Si*, Philosophical Magazine B **79**, 1715 (1999).

[4] C. Kittel, *Interpretation of the Thermal Conductivity of Glasses*, Physical Review **75**, 972 (1949).

[5] R. C. Zeller and R. O. Pohl, *Thermal Conductivity and Specific Heat of Noncrystalline Solids*, Physical Review B **4**, 2029 (1971).

[6] A. Eucken, *Über Die Temperaturabhängigkeit Der Wärmeleitfähigkeit Fester Nichtmetalle*, Annalen Der Physik **339**, 185 (1911).

[7] J. Moon, *Examining Normal Modes as Fundamental Heat Carriers in Amorphous Solids: The Case of Amorphous Silicon*, Journal of Applied Physics **130**, 055101 (2021).

[8] F. DeAngelis, M. G. Muraleedharan, J. Moon, H. R. Seyf, A. J. Minnich, A. J. H. McGaughey, and A. Henry, *Thermal Transport in Disordered Materials*, Nanoscale and Microscale Thermophysical Engineering (2018).

[9] F. Sette, *Dynamics of Glasses and Glass-Forming Liquids Studied by Inelastic X-Ray Scattering*, Science **280**, 1550 (1998).

[10] P. Benassi, M. Krisch, C. Masciovecchio, V. Mazzacurati, G. Monaco, G. Ruocco, F. Sette, and R. Verbeni, *Evidence of High Frequency Propagating Modes in Vitreous Silica*, Physical Review Letters **77**, 3835 (1996).

[11] F. Sette, G. Ruocco, M. Krisch, U. Bergmann, C. Masciovecchio, V. Mazzacurati, G. Signorelli, and R. Verbeni, *Collective Dynamics in Water by High Energy Resolution Inelastic X-Ray Scattering*, Physical Review Letters **75**, 850 (1995).

[12] G. Baldi, V. M. Giordano, B. Ruta, R. Dal Maschio, A. Fontana, and G. Monaco, *Anharmonic Damping of Terahertz Acoustic Waves in a Network Glass and Its Effect on the Density of Vibrational States*, Physical Review Letters **112**, (2014).

[13] J. Moon, R. P. Hermann, M. E. Manley, A. Alatas, A. H. Said, and A. J. Minnich, *Thermal Acoustic Excitations with Atomic-Scale Wavelengths in Amorphous Silicon*, Physical Review Materials **3**, 065601 (2019).

[14] E. A. A. Pogna, C. Rodríguez-Tinoco, M. Krisch, J. Rodríguez-Viejo, and T. Scopigno, *Acoustic-like Dynamics of Amorphous Drugs in the THz Regime*, Scientific Reports **3**, (2013).

[15] C. Masciovecchio, A. Mermet, G. Ruocco, and F. Sette, *Experimental Evidence of the Acousticlike Character of the High Frequency Excitations in Glasses*, Physical Review Letters **85**, 1266 (2000).

[16] C. Masciovecchio, G. Monaco, G. Ruocco, F. Sette, A. Cunsolo, M. Krisch, A. Mermet, M. Soltwisch, and R. Verbeni, *High Frequency Dynamics of Glass Forming Liquids at the Glass Transition*, Physical Review Letters **80**, 544 (1998).

[17] T. Kim, J. Moon, and A. J. Minnich, *Origin of Micrometer-Scale Propagation Lengths of Heat-Carrying Acoustic Excitations in Amorphous Silicon*, Phys. Rev. Materials **5**, 065602 (2021).

[18] B. C. Daly, K. Kang, Y. Wang, and D. G. Cahill, *Picosecond Ultrasonic Measurements of Attenuation of Longitudinal Acoustic Phonons in Silicon*, Physical Review B **80**, (2009).

[19] D. B. Hondongwa, B. C. Daly, T. B. Norris, B. Yan, J. Yang, and S. Guha, *Ultrasonic Attenuation in Amorphous Silicon at 50 and 100 GHz*, Physical Review B **83**, (2011).

[20] J. Moon, B. Latour, and A. J. Minnich, *Propagating Elastic Vibrations Dominate Thermal Conduction in Amorphous Silicon*, Physical Review B **97**, (2018).

[21] Y. M. Beltukov, C. Fusco, D. A. Parshin, and A. Tanguy, *Boson Peak and Ioffe-Regel Criterion in Amorphous Siliconlike Materials: The Effect of Bond Directionality*, Physical Review E **93**, (2016).

[22] H. Shintani and H. Tanaka, *Universal Link between the Boson Peak and Transverse Phonons in Glass*, Nature Materials **7**, 870 (2008).

[23] G. Monaco and S. Mossa, *Anomalous Properties of the Acoustic Excitations in Glasses on the Mesoscopic Length Scale*, Proceedings of the National Academy of Sciences **106**, 16907 (2009).

[24] G. Ruocco and F. Sette, *High-Frequency Vibrational Dynamics in Glasses*, Journal of Physics: Condensed Matter **13**, 9141 (2001).

[25] G. Ruocco, F. Sette, R. Di Leonardo, D. Fioretto, M. Krisch, M. Lorenzen, C. Masciovecchio, G. Monaco, F. Pignon, and T. Scopigno, *Nondynamic Origin of the High-Frequency Acoustic Attenuation in Glasses*, Physical Review Letters **83**, 5583 (1999).

[26] G. Monaco and V. M. Giordano, *Breakdown of the Debye Approximation for the Acoustic Modes with Nanometric Wavelengths in Glasses*, Proceedings of the National Academy of Sciences **106**, 3659 (2009).

[27] S. Gelin, H. Tanaka, and A. Lemaître, *Anomalous Phonon Scattering and Elastic Correlations in Amorphous Solids*, Nature Materials **15**, 1177 (2016).

[28] C. Masciovecchio, G. Baldi, S. Caponi, L. Comez, S. Di Fonzo, D. Fioretto, A. Fontana, A. Gessini, S. C. Santucci, F. Sette, G. Viliani, P. Vilmercati, and G. Ruocco, *Evidence for a Crossover in the Frequency Dependence of the Acoustic Attenuation in Vitreous Silica*, Physical Review Letters **97**, (2006).

[29] G. Baldi, V. M. Giordano, G. Monaco, and B. Ruta, *Sound Attenuation at Terahertz Frequencies and the Boson Peak of Vitreous Silica*, Physical Review Letters **104**, (2010).

[30] D. Fioretto, U. Buchenau, L. Comez, A. Sokolov, C. Masciovecchio, A. Mermet, G. Ruocco, F. Sette, L. Willner, and B. Frick, *High-Frequency Dynamics of Glass-Forming Polybutadiene*, Physical Review E **59**, 4470 (1999).

[31] W. Schirmacher, G. Ruocco, and T. Scopigno, *Acoustic Attenuation in Glasses and Its Relation with the Boson Peak*, Phys. Rev. Lett. **98**, 025501 (2007).

[32] Y. M. Beltukov, V. I. Kozub, and D. A. Parshin, *Ioffe-Regel Criterion and Diffusion of Vibrations in Random Lattices*, Phys. Rev. B **87**, 134203 (2013).

[33] T. Scopigno, J.-B. Suck, R. Angelini, F. Albergamo, and G. Ruocco, *High-Frequency Dynamics in*

*Metallic Glasses*, Physical Review Letters **96**, (2006).

[34] P. Bruna, G. Baldi, E. Pineda, J. Serrano, J. B. Suck, D. Crespo, and G. Monaco, *Communication: Are Metallic Glasses Different from Other Glasses? A Closer Look at Their High Frequency Dynamics*, The Journal of Chemical Physics **135**, 101101 (2011).

[35] A. De Francesco, U. Bafile, A. Cunsolo, L. Scaccia, and E. Guarini, *Searching for a Second Excitation in the Inelastic Neutron Scattering Spectrum of a Liquid Metal: A Bayesian Analysis*, Sci Rep **11**, 13974 (2021).

[36] N. Jakse, A. Nassour, and A. Pasturel, *Structural and Dynamic Origin of the Boson Peak in a Cu-Zr Metallic Glass*, Phys. Rev. B **85**, 174201 (2012).

[37] D. Crespo, P. Bruna, A. Valles, and E. Pineda, *Phonon Dispersion Relation of Metallic Glasses*, Physical Review B **94**, (2016).

[38] G. Baldi, A. Fontana, and G. Monaco, *Vibrational Dynamics of Non-Crystalline Solids*, 47 (n.d.).

[39] S. Plimpton, *Fast Parallel Algorithms for Short-Range Molecular Dynamics*, Journal of Computational Physics **117**, 1 (1995).

[40] Y. Q. Cheng and E. Ma, *Atomic-Level Structure and Structure–Property Relationship in Metallic Glasses*, Progress in Materials Science **56**, 379 (2011).

[41] D. Srolovitz, K. Maeda, V. Vitek, and T. Egami, *Structural Defects in Amorphous Solids Statistical Analysis of a Computer Model*, Philosophical Magazine A **44**, 847 (1981).

[42] H. Mizuno and A. Ikeda, *Phonon Transport and Vibrational Excitations in Amorphous Solids*, Physical Review E **98**, 062612 (2018).

[43] P. Sun, J. B. Hastings, D. Ishikawa, A. Q. R. Baron, and G. Monaco, *Two-Component Dynamics and the Liquidlike to Gaslike Crossover in Supercritical Water*, Phys. Rev. Lett. **125**, 256001 (2020).

[44] L. S. Ornstein and F. Zernike, *Accidental Deviations of Density and Opalescence at the Critical Point of a Single Substance*, Royal Netherlands Academy of Arts and Sciences Proceedings **17**, 793 (1914).

[45] C. W. Ryu, W. Dmowski, K. F. Kelton, G. W. Lee, E. S. Park, J. R. Morris, and T. Egami, *Curie-Weiss Behavior of Liquid Structure and Ideal Glass State*, Scientific Reports **9**, 18579 (2019).

[46] T. Egami, *Local Density Correlations in Liquids*, Frontiers in Physics **8**, 50 (2020).

[47] C. W. Ryu and T. Egami, *Origin of Liquid Fragility*, Physical Review E **102**, 042615 (2020).

[48] J. Moon and T. Egami, *Enhancing Elastic Properties of Single Element Amorphous Solids through Long-Range Interactions*, Appl. Phys. Lett. **119**, 051901 (2021).

[49] P. G. Klemens, *The Scattering of Low-Frequency Lattice Waves by Static Imperfections*, Proc. Phys. Soc. A **68**, 1113 (1955).

[50] P. Carruthers, *Scattering of Phonons by Elastic Strain Fields and the Thermal Resistance of Dislocations*, Phys. Rev. **114**, 995 (1959).

[51] M. Arrigoni, J. Carrete, N. Mingo, and G. K. H. Madsen, *First-Principles Quantitative Prediction of the Lattice Thermal Conductivity in Random Semiconductor Alloys: The Role of Force-*

Constant Disorder, Phys. Rev. B **98**, 115205 (2018).

[52] F. H. Stillinger, *A Topographic View of Supercooled Liquids and Glass Formation*, Science **267**, 1935 (1995).

[53] S. Sastry, P. G. Debenedetti, and F. H. Stillinger, *Signatures of Distinct Dynamical Regimes in the Energy Landscape of a Glass-Forming Liquid*, Nature **393**, 554 (1998).

[54] Y. Fan, T. Iwashita, and T. Egami, *Energy Landscape-Driven Non-Equilibrium Evolution of Inherent Structure in Disordered Material*, Nat Commun **8**, 15417 (2017).

[55] N. Mousseau and G. T. Barkema, *Traveling through Potential Energy Landscapes of Disordered Materials: The Activation-Relaxation Technique*, Physical Review E **57**, 2419 (1998).

[56] H. Kallel, N. Mousseau, and F. Schiettekatte, *Evolution of the Potential-Energy Surface of Amorphous Silicon*, Physical Review Letters **105**, 045503 (2010).

[57] P. W. Anderson, B. I. Halperin, and C. M. Varma, *Anomalous Low-Temperature Thermal Properties of Glasses and Spin Glasses*, The Philosophical Magazine: A Journal of Theoretical Experimental and Applied Physics **25**, 1 (1972).

[58] W. A. Phillips, *Tunneling States in Amorphous Solids*, Journal of Low Temperature Physics **7**, 351 (1972).

[59] J. Jäckle, *On the Ultrasonic Attenuation in Glasses at Low Temperatures*, Zeitschrift Für Physik A Hadrons and Nuclei **257**, 212 (1972).

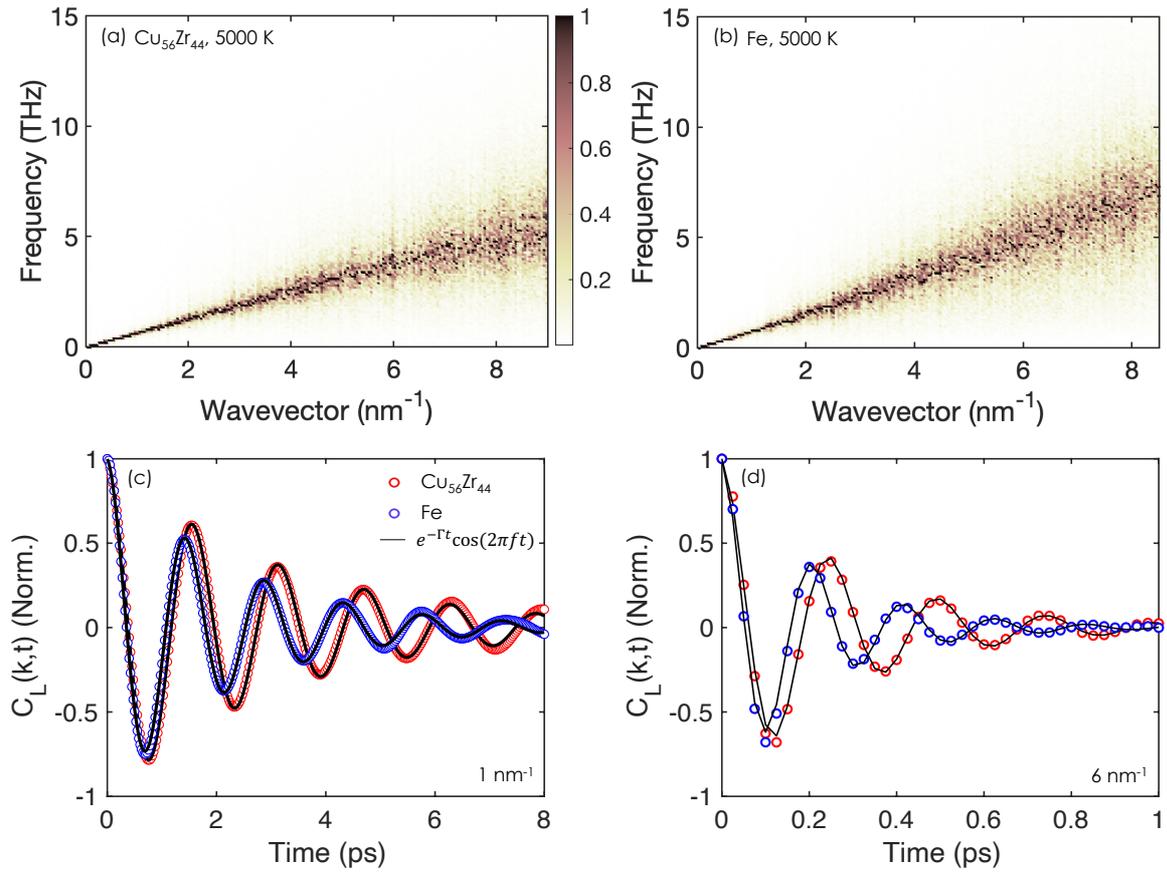

Fig 1. Longitudinal vibrational excitations dispersion for (a) liquid $Cu_{56}Zr_{44}$ at 5000 K and (b) liquid Fe at 5000 K. For better visualizations, each spectra is normalized by its maximum at each wavevector. Select time-resolved longitudinal current-current correlations at (c) 1 nm$^{-1}$ and (d) 6 nm$^{-1}$ at 5000 K. Red circles and blue circles are for liquid $Cu_{56}Zr_{44}$ and liquid Fe, respectively. Solid black lines are the fitted curves to extract the excitation frequency and attenuation. Excellent fits are observed.

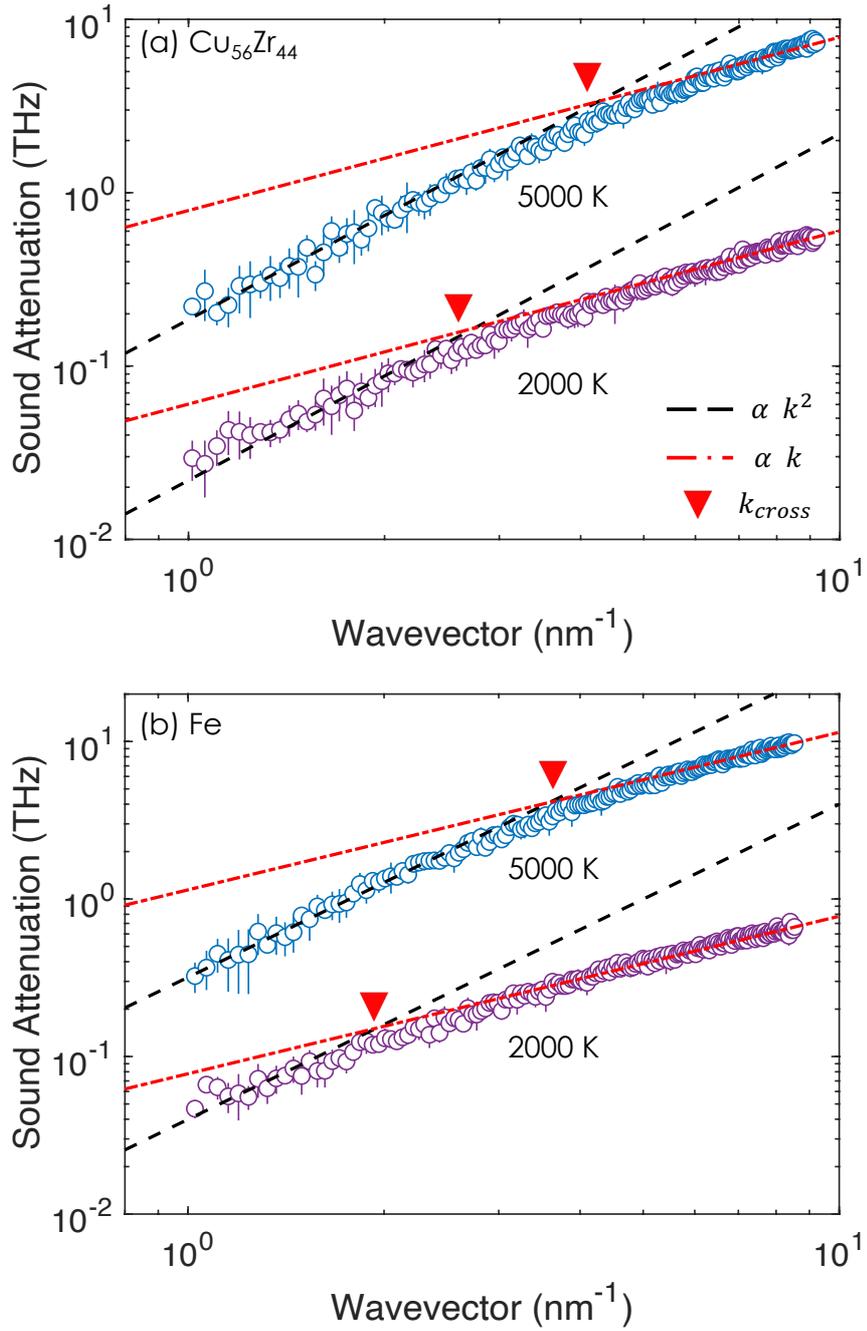

Fig 2. Sound attenuation rates for (a) liquid $Cu_{56}Zr_{44}$ and (b) liquid Fe at 2000 K and 5000 K. For ease of visualizations, the depicted attenuation rates for 2000 K have been scaled by a factor of 0.1. Black dashed lines and red dash-dotted lines represent quadratic and linear dependence in wavevector, respectively. Red solid triangle denotes the crossover wavevector where the sound attenuation changes from quadratic to linear dependence. Temperature dependence in the crossover wavevector is clearly shown.

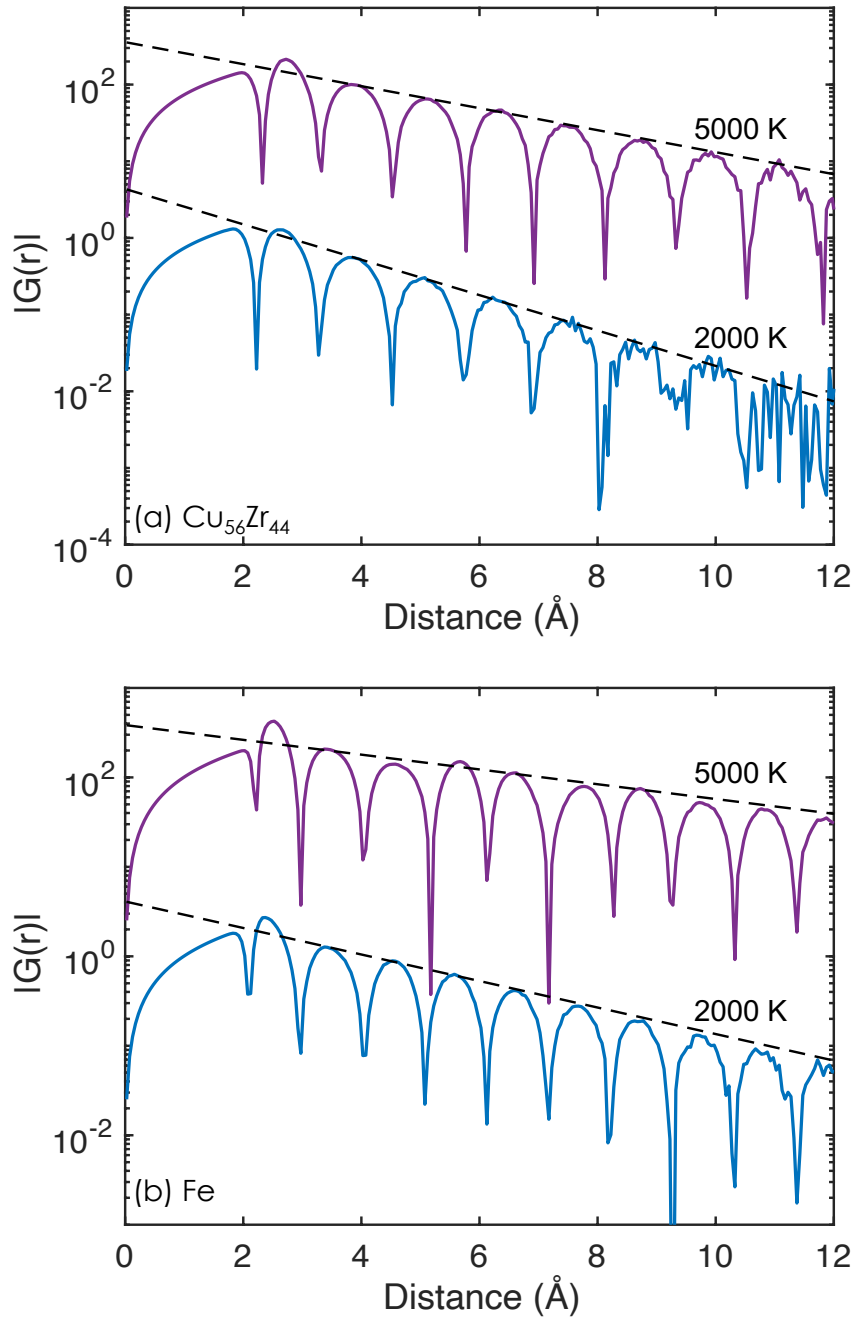

Fig 3. Magnitude of reduced pair distribution functions for (a) liquid $Cu_{56}Zr_{44}$ and (b) liquid Fe at 2000 K and 5000 K. Black dashed lines are fitted lines to extract the structural coherence lengths characterizing the medium range order. The reduced pair distributions at 2000 K are scaled by a factor of 100. In both systems, a clear temperature dependence in structural coherence lengths is shown.

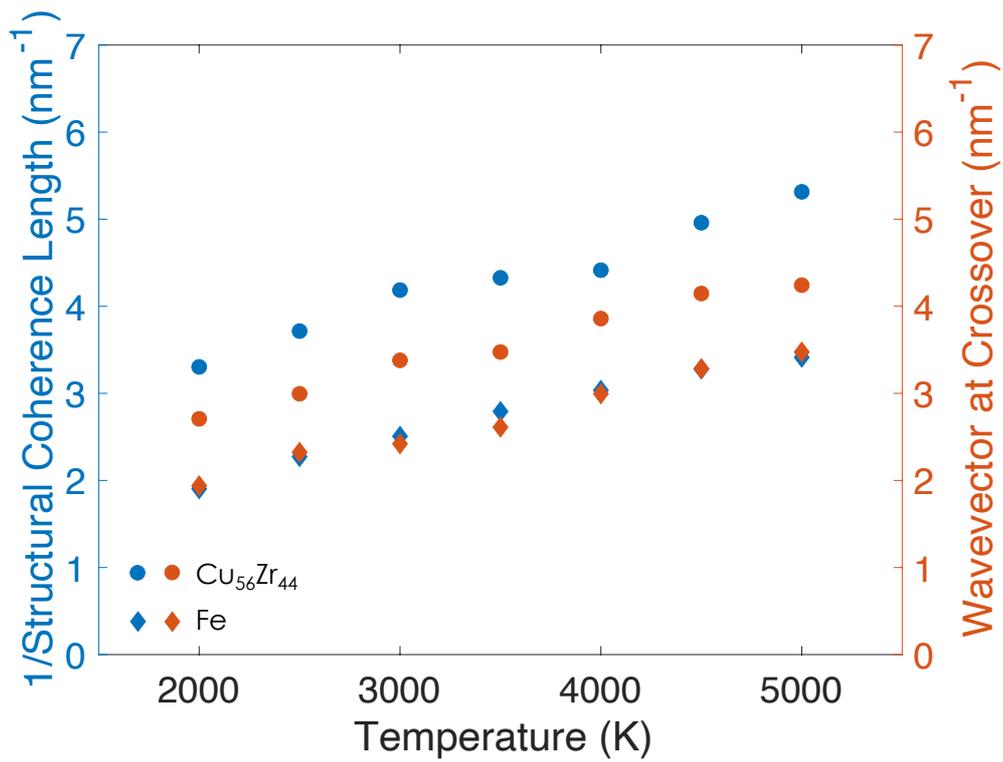

Fig 4. Temperature dependent wavevector (orange) at the power law crossover in attenuation rates and inverse of structural coherence length (blue) for liquid $Cu_{56}Zr_{44}$ (solid circles) and liquid Fe (solid diamonds). We see that the temperature dependence of these two quantities are nearly identical, signifying that the linear attenuation rate regime is strongly influenced by the medium range order.

**SUPPLEMENTARY MATERIAL FOR**

**"Structural effect on phonon attenuation in metallic liquids and glasses"**

Jaeyun Moon[1] and Takeshi Egami[1,2,3]

[1]*Materials Science and Technology Division,*
*Oak Ridge National Laboratory, Oak Ridge, Tennessee 37831, USA*
[2]*Department of Materials Science and Engineering,*
*University of Tennessee, Knoxville, Tennessee 37996, USA*
[3]*Department of Physics and Astronomy,*
*University of Tennessee, Knoxville, Tennessee 37996, USA*

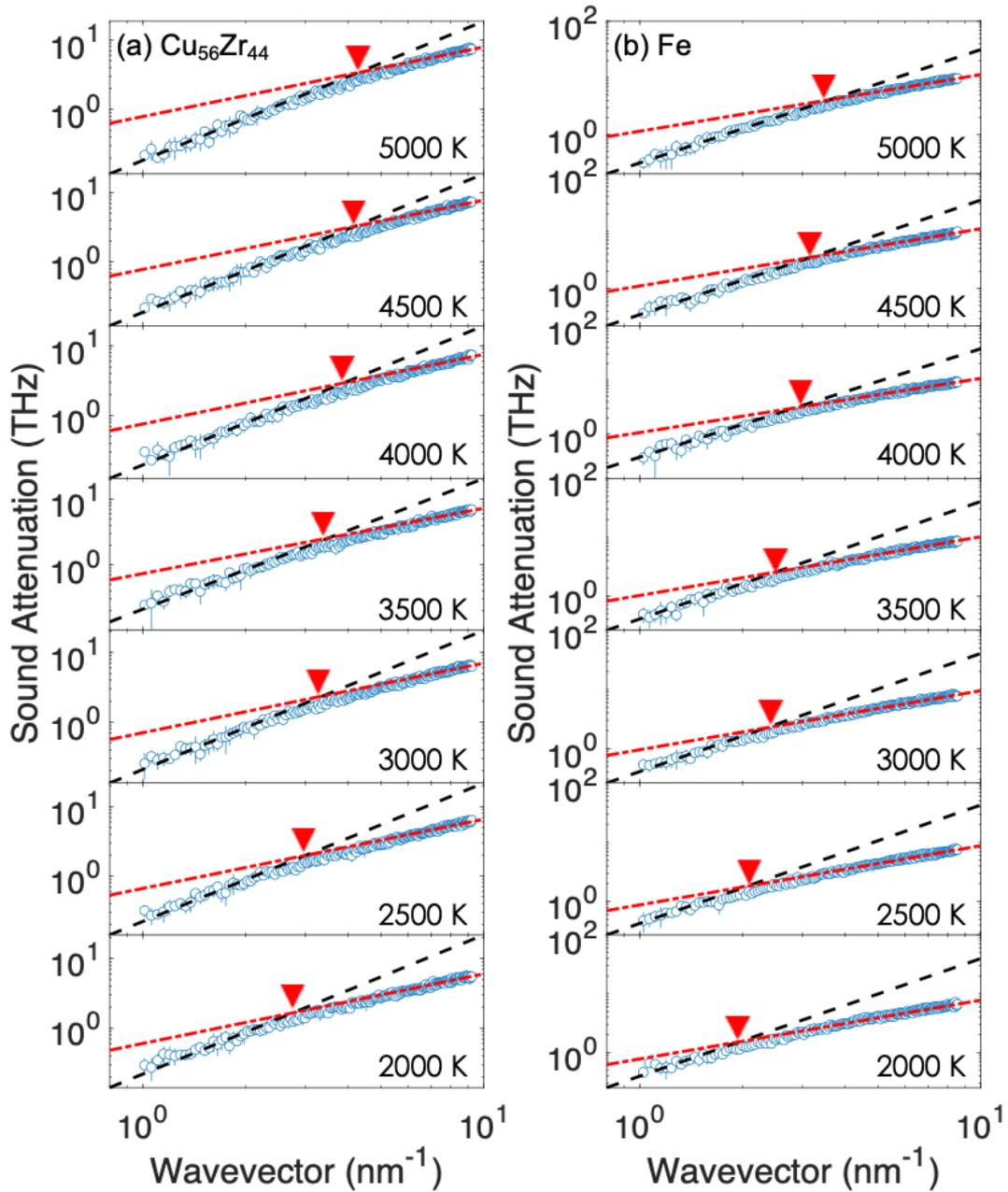

Fig S1. Sound attenuation rates for (a) liquid $Cu_{56}Zr_{44}$ and (b) liquid Fe at various temperatures from 5000 K to 2000 K. Markers and lines are identical to Fig. 2 in the main manuscript. A clear temperature dependence in the wavevector at which the power law changes from quadratic to linear dependence is shown for both metallic systems.

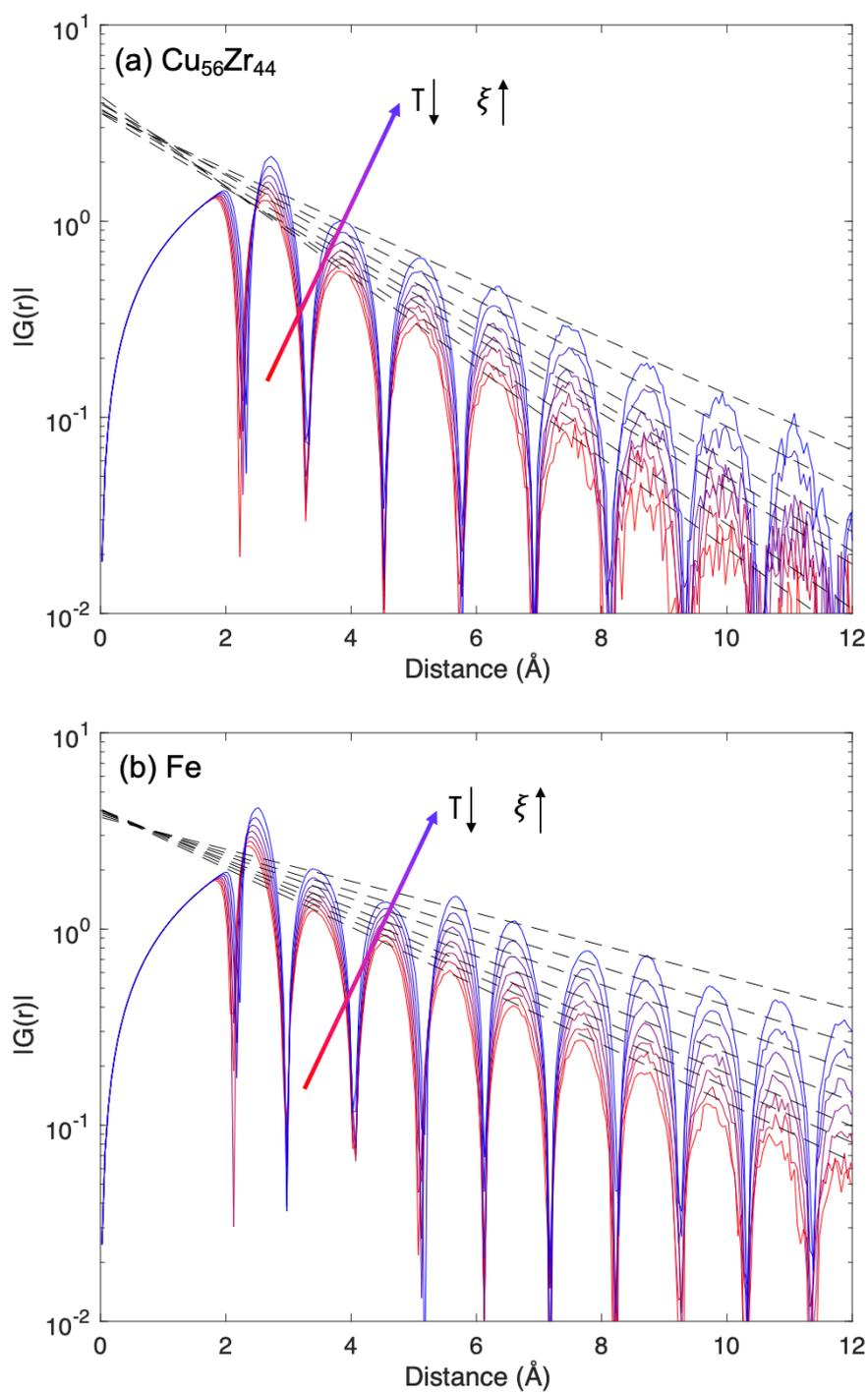

Fig S2. Magnitude of reduced pair distribution functions for (a) liquid $Cu_{56}Zr_{44}$ and (b) liquid Fe at various temperatures from 5000 K to 2000 K. Red curves denotes 5000 K and blue curves denotes 2000 K. Black dashed lines are the exponential decay fit as mentioned in the main manuscript. Clear changes in the structural coherence lengths with temperature are demonstrated.